\documentclass[aps,pra,preprint,superscriptaddress,showpacs]{revtex4-1}

\usepackage[latin1]{inputenc}
\usepackage[english]{babel}
\usepackage{amsfonts,amssymb,amsmath,amscd,mathtools,slashed}
\usepackage[amsthm,thmmarks,amsmath]{ntheorem}
\usepackage{enumerate}
\usepackage{color}
\usepackage{verbatim}

\usepackage{graphicx}
\usepackage{tikz}
\usepackage{mathrsfs}

\newtheorem{thm}{Theorem}

\theoremstyle{definition}
\newtheorem{defn}[thm]{Definition}

\newtheorem{example}[thm]{Example}

\newcommand{\abs}[1]{\left\vert {#1} \right\vert}

\newcommand{\scal}[1]{\left< {#1} \right>}

\newcommand{\sN}{{\mathbb N}}

\newcommand{\sR}{{\mathbb R}}
\newcommand{\sC}{{\mathbb C}}

\newcommand{\A}{\mathcal{A}}

\newcommand{\C}{\mathcal{C}}

\newcommand{\F}{\mathcal{F}}

\renewcommand{\H}{\mathcal{H}}

\newcommand{\M}{\mathcal{M}}

\newcommand{\K}{\mathcal{K}}

\newcommand{\CC}{\mathcal{C}}

\DeclareMathOperator{\Span}{span}

\newcommand{\dt}{\partial}
\newcommand{\DD}{\mathcal{D}}           
\newcommand{\ox}{\otimes}                  
\newcommand{\Dslash}{{\DD \mkern-11.5mu/\,}} 

\newcommand{\x}{\times}
\newcommand{\vc}{\vcentcolon =}
\newcommand{\cv}{= \vcentcolon}

\newcommand{\<}{\left\langle}		
\renewcommand{\>}{\right\rangle}		
\newcommand{\wA}{\mathbb{A}}
\renewcommand{\bar}{\overline}		
\newcommand{\Aslash}{{A \mkern-9mu/\,}} 
\newcommand{\DA}{\DD_{\mathbb{A}}}	

\renewcommand{\M}{M}
\newcommand{\Hs}{\H_{\text{sim}}}

\begin{document}


\title{The noncommutative geometry of \textit{Zitterbewegung}}

\author{Micha{\l} Eckstein}
\email{michal.eckstein@uj.edu.pl}
\affiliation{Faculty of Physics, Astronomy and Applied Computer Science, Jagiellonian University, ul. prof. Stanis{\l}awa {\L}ojasiewicza 11, 30-348 Krak\'ow, Poland}
\affiliation{Copernicus Center for Interdisciplinary Studies, ul. S{\l}awkowska 17, 31-016 Krak\'ow, Poland}

\author{Nicolas Franco}
\email{nicolas.franco@math.unamur.be}
\affiliation{Namur Center for Complex Systems (naXys) \& Department of Mathematics, University of Namur, Rue de Bruxelles 61, 5000 Namur, Belgium}
\affiliation{Copernicus Center for Interdisciplinary Studies, ul. S{\l}awkowska 17, 31-016 Krak\'ow, Poland}

\author{Tomasz Miller}
\email{t.miller@mini.pw.edu.pl}
\affiliation{Faculty of Mathematics and Information Science, Warsaw University of Technology, ul. Koszykowa 75, 00-662 Warsaw, Poland}
\affiliation{Copernicus Center for Interdisciplinary Studies, ul. S{\l}awkowska 17, 31-016 Krak\'ow, Poland}

\date{\today}

\begin{abstract}
Based on the mathematics of noncommutative geometry, we model a `classical' Dirac fermion propagating in a curved spacetime. We demonstrate that the inherent causal structure of the model encodes the possibility of \textit{Zitterbewegung} -- the `trembling motion' of the fermion. We recover the well-known frequency of \textit{Zitterbewegung} as the highest possible speed of change in the fermion's `internal space'. Furthermore, we show that the latter does not change in the presence of an external electromagnetic field and derive its explicit analogue when the mass parameter is promoted to a Higgs-like field. We discuss a table-top experiment in the domain of quantum simulation to test the predictions of the model and outline the consequences of our model for quantum gauge theories.
\end{abstract}


\maketitle


The set-up of the Dirac equation \cite{DiracBook} was a prodigious step towards the unification of quantum and relativistic principles. Surprisingly enough, the differential operator engaged in the Dirac equation turned out to play a pivotal role in differential geometry \cite{Friedrich}. Moreover, it lies at the heart of Connes' theory of noncommutative geometry \cite{C94,MC08}, which extends such classical notions as differentiation, distance \cite{C94} and 
causality \cite{CQG2013} to an abstract algebraic setting. Nowadays, noncommutative geometry centred around the concept of the Dirac operator provides a compelling framework for the study of fundamental interactions \cite{CC96}, yielding concrete testable predictions in the domain of elementary particles \cite{Almost1,WalterBook} and gravitational physics \cite{MarcolliCosmSurv,MairiJCAP}.

Building upon Connes' ideas \cite{ConnesZitter} we model a single massive Dirac fermion with the help of an almost commutative spacetime. The latter turns out to provide a geometric description of one of the peculiarities of the Dirac theory -- the \emph{Zitterbewegung} \cite{ZitterSchroedinger}. We show that the causal structure of the almost commutative spacetime at hand puts an explicit bound on the frequency of the `trembling motion' of a single Dirac fermion propagating in a possibly curved spacetime and interacting with the electromagnetic and a Higgs-like scalar field. Next, we explain how the concept of quantum simulation  \cite{QuantumSimulation} can be promoted to emulate almost commutative spacetimes, thus opening the door to a direct experimental test of Connes' theory.

\section*{Zitterbewegung}
The Dirac equation,
\begin{align}\label{Dirac}
\left(i \hbar \gamma^{\mu} \dt_{\mu} - m c \right) \psi = 0,
\end{align}
reveals a number of striking facts about the nature of massive fermions. One of them concerns the velocity operator in the Heisenberg picture, $\hat{v}_k(t) \vc \tfrac{\dt \hat{x}_k(t)}{\dt t}$, which turns out to have eigenvalues $\pm c$ for all moments of time \cite{Thaller}. This suggests that the instantaneous velocity of a \emph{massive} fermion is always $\pm c$, which seems paradoxical. A more detailed analysis unveils that both the velocity and the position operators in the Heisenberg picture have a part that oscillates in time, hence the name  \textit{Zitterbewegung} --- `the trembling motion' --- coined by Schr\"odinger  \cite{ZitterSchroedinger}.

For an initial state with vanishing average momentum, the expectation value of the position operator oscillates 
 with the period \cite{ZitterSchroedinger,ZitterBEC,ZitterPark}
\begin{align}\label{zitter_period}
T_{\text{ZB}} = \pi \hbar/(m c^2).
\end{align}

The original explanation of the  mechanism behind \textit{Zitterbewegung} given by Schr\"odinger \cite{ZitterSchroedinger}, and refined by several authors \cite{Thaller,ZitterHuang,ZitterPark}, relates it to the interference between the positive and negative energy parts of the Dirac wave packet. Indeed, if the initial state $\psi$ is a purely positive (or negative) energy state then the expectation value $\scal{\psi,\hat{x}_k(t) \psi}$ does not exhibit oscillations \cite{Thaller,ZitterSchroedinger}.

The tangibility of \textit{Zitterbewegung} for actual fermions being \textit{par force} positive-energetic excitations of a quantum field is generally questioned \cite{ZitterQFT} (see however \cite{ZitterCosmo}). On the other hand, the realness of this effect has been confirmed in various Dirac-like systems \cite{ZitterBEC,ZitterNature,ZitterPhotonics} and was found responsible for the appearance of a minimal conductivity and a sub-Poissonian shot noise in graphene \cite{ZitterGraphene1,ZitterGraphene2}.

There exists an alternative viewpoint on \textit{Zitterbewegung} (known also under the name of `chiral oscillations' \cite{ChiralZitter,ChiralZitter2}) relating it to the spin components of the fermionic wave function \cite{Hestenes1990,Hestenes2003,Hestenes2010,PenroseBook,PenroseSpin}. 
Any Dirac spinor can be uniquely decomposed as a sum of two Weyl spinors $\psi = \psi_+ + \psi_-$, where $\psi_{\pm} = P_{\pm} \psi$, with $P_{\pm} = (1 \pm \gamma^5)/2$ \cite{PenroseSpin}. The Weyl spinors $\psi_{\pm}$, being eigenstates of the chirality operator $\gamma^5$ with $\gamma^5 \psi_{\pm} = \pm \psi_{\pm}$, have opposite chirality. By acting with the projector $P_{\pm}$ on Equation \eqref{Dirac} we obtain
\begin{align}\label{Dirac_Penrose}
i \hbar \gamma^{\mu} \dt_{\mu} \psi_{\pm} =  m c \psi_{\mp},
\end{align}
which can be seen as two coupled equations for Weyl spinors $\psi_{\pm}$, one acting as a source of the other \cite{PenroseSpin}. Since Weyl spinors are massless, they move with the speed of light and one obtains a `zigzag picture' of a massive fermion \cite[Figure 25.1]{PenroseBook}. In the fermion's rest frame, the period of oscillations between the two eigenstates of chirality precisely equals \eqref{zitter_period} \cite{Hestenes2010}.

The two points of view on \textit{Zitterbewegung} are closely related in the Dirac wave packet formalism \cite{ChiralZitter2}. In particular, purely positive/negative energy solutions to the Dirac equation also do not exhibit chiral oscillations \cite{ChiralZitter}.

Hestenes contended \cite{Hestenes1990,Hestenes2003,Hestenes2010} that the `chiral' interpretation of \textit{Zitterbewegung} is more natural, as the origin of the effect resides in the geometry of spacetime. In the present Letter we support this claim, although we argue that the very notion of geometry needs to be refined.

\section*{\label{NCG}Noncommutative geometry}
The basic objects of noncommutative geometry \`a la Connes \cite{C94} are spectral triples $(\A,\H,\DD)$ consisting of a (dense subalgebra of a) $C^*$-algebra $\A$, a Hilbert space $\H$ with a faithful representation of $\A$ and an unbounded self-adjoint operator $\DD$ acting on $\H$. The original framework was designed to describe spaces of Euclidean signature and has recently been extended to encompass the Lorentzian ones \cite{Stro,Pas,F4,Verch11,Rennie12,F5}. In the latter case, the main conceptual change consists in endowed the Hilbert space $\H$ with an indefinite inner product, turning it into a Krein space $\K$ \cite{Bog} --- a vector space equipped with an indefinite non-degenerate inner product --- and in requiring $\DD$ to be self-adjoint with respect to the indefinite inner product.

\begin{example}\label{ex:manifold}
Let $M$ be a globally hyperbolic spacetime with a spin structure, then \linebreak $(\A_M,\K_M,\Dslash)$ is a Lorentzian spectral triple, with $\A_M = C^{\infty}_c(M)$ -- the algebra of smooth compactly supported functions on $M$, $\K_M = L^2(M,S)$ -- the space of square summable sections of the spinor bundle $S$ over $M$, and $\Dslash = - i \gamma^{\mu} \nabla^S_{\mu}$ -- the (curved) Dirac operator associated with $S$. If $\<\cdot,\cdot\>$ stands for a Hermitian inner product on $L^2(M,S)$, then the indefinite inner product on $\K_M$ can be defined as $(\cdot,\cdot) = \<\cdot \gamma^0,\cdot\>$ with $\gamma^0$ being the Hermitian first flat gamma matrix.
\end{example}

Even-dimensional spacetimes induce an additional structure on the associated spectral triple -- a chirality operator $\gamma_M = \gamma^5$, splitting the space $\K_M = \K_M^+ \oplus \K_M^-$, with $\gamma_M \psi_\pm = \pm \psi_{\pm}$ for $\psi_{\pm} \in \K_M^{\pm}$.

The most important examples of spectral triples from the viewpoint of physical applications are the \emph{almost commutative geometries} \cite{WalterBook,SIGMA2014}:

\begin{example}\label{ex:almost}
Let $(\A_F,\H_F,\DD_F)$ be a \emph{finite} spectral triple, i.e. $\A_F = M_n(\sC)$, $\H_F = \sC^n$ and $\DD_F = \DD_F^{\dagger} \in M_n(\sC)$ for some $n \in \sN$, and let $(\A_M,\K_M,\Dslash)$ be as in Example \ref{ex:manifold} with $\dim M$ even. Then, the data $(\A,\K,\DD)$ = $(\A_M \ox \A_F$, $\K_M \ox \H_F$, $\Dslash \ox 1 + i\gamma_M \ox \DD_F)$ form a Lorentzian spectral triple. (The $i$ factor is used in order to keep the spectral triple Lorentzian with the convention of signature $(+,-,-,-)$ used in this paper.)
\end{example}

\section*{\label{almost}Causality and almost commutative spacetimes}
Causality is one of the most fundamental principles underlying physical theories. Within Einstein's theory it is defined as a partial order relation on the set of events: $p \preceq q$ means that $q$ lies in the future of $p$. However, noncommutative spaces (not only in the framework of spectral triples) typically admit only a global description and the very notion of an event does not make sense. This raises the question: what is the scene for causal relations and what is an operational meaning of a `noncommutative spacetime'?

In \cite{CQG2013} we put forward the idea that a `noncommutative spacetime' ought to be understood as the space of (pure) states of a, possibly noncommutative, $C^*$-algebra. The motivation behind this step is twofold: Firstly, if the algebra at hand is of the form $\A_M = C^{\infty}_c(M)$ (see Example \ref{ex:manifold}), then its pure states $P(\A_M)$ are in one-to-one correspondence with the events in $M$ \cite{CQG2013}. 
Secondly, $C^*$-algebras provide a unified framework for an operational formulation of both classical and quantum physics \cite{Haag,Strocchi,Keyl}. In this context, states on an abstract $C^*$-algebra of observables can be understood as the actual states of a given physical system. For instance, if $\A_F = M_n(\sC)$ (see Example \ref{ex:almost}), then pure states in $P(\A_F)$ are precisely the $n$-qubits, whereas the mixed states $S(\A_F)$ correspond to the density matrices \cite{Keyl}.

In \cite{CQG2013} we have shown that such a `noncommutative spacetime' admits a sensible notion of a causal structure associated with a Lorentzian spectral triple. To this end, one has to identify a specific subset $\C$ of the algebra $\A$ named the `causal cone'. Concretely, $\C$ is the cone of all Hermitian elements $a$ of a preferred unitisation $\wA \supseteq \A$ respecting $\forall \; \phi~\in~\K,\; (\phi,[\DD,a] \phi) \geq 0$, with the additional condition $\overline{\Span_{{\sC}}(\CC)}=\overline{\wA}$ (see \cite{CQG2013,SIGMA2014,PROC2015} for the details). 

\begin{defn}\label{def:causcone}
Let $(\A,\K,\DD)$ be a Lorentzian spectral triple and let $\C$ be the causal cone. We say that two states $\omega, \eta \in S(\A)$ are causally related with $\omega\preceq\eta$ iff $\forall\, a\in\C \; \omega(a)\leq\eta(a)$.
\end{defn}


Definition \ref{def:causcone} designates a partial order relation on $S(\A)$ and is strongly motivated by the following result \cite{CQG2013}.

\begin{thm}\label{thm:reconstruction}
Let $(\A_M,\K_M,\Dslash)$ be a Lorentzian spectral triple constructed from a globally hyperbolic 
spin manifold $M$. Then, two pure states $\phi_p, \phi_q \in P(\A_M) \cong M$ are causally related with $\phi_p \preceq \phi_q$ if and only if $p \preceq q$ in $M$.
\end{thm}

It justifies calling the space $P(\A)$ equipped with the partial order $\preceq$ a `noncommutative spacetime'. Note also that Definition \ref{def:causcone} determines a causal order on the significantly larger space $S(\A)$ of mixed states. We shall exploit this fact when discussing the experimental setup.

The space of pure states of an almost commutative geometry has a particularly pellucid physical interpretation: Let $\A = \A_M \ox \A_F$ as in Example \ref{ex:almost}, then $P(\A) \simeq M \times \F$ for some finite dimensional space $\F$. In other words, all pure states on $\A$ are separable \cite{Kadison}. Hence, the space of pure states of an \emph{almost commutative spacetime} is the Cartesian product of the spacetime $M$ and an `inner' space of states of the model. Almost commutative spacetimes have a desired property \cite{PROC2015}:

\begin{thm}\label{thm:noviolation}
Let $P(\A)$ be an almost commutative spacetime. If $(p,\xi), (q,\chi) \in P(\A)$ are such that $(p,\xi) \preceq (q,\chi)$, then $p \preceq q$ in the spacetime $\M$.
\end{thm}

This result attests that in almost commutative spacetimes, Einstein's causality in the spacetime component is not violated. On the other hand, in \cite{SIGMA2014} and \cite{JGP2015} we discovered that the extended causal structure imposes highly non-trivial restrictions on the evolution also in the `inner' space of the model. We shall now apply the mathematical results obtained in \cite{JGP2015} to lift the veil on the nature of \textit{Zitterbewegung}.


\section*{\label{sec:model}Modelling a `classical' Dirac fermion}
Let $M$ be a globally hyperbolic spacetime of dimension 2 or 4. We associate to it a Lorentzian spectral triple $(\A_M,\K_M,\Dslash)$ in a canonical way (see Example \ref{ex:manifold}). As a finite spectral triple we take $\A_F = \sC \oplus \sC$, $\H_F = \sC^2$ and $\DD_F = \left(\begin{smallmatrix}0 & \mu\\ \bar{\mu} & 0\end{smallmatrix}\right)$, for some $\mu \in \sC \setminus \{0\}$. The product triple thus reads: $\A = C^{\infty}_c(M) \oplus C^{\infty}_c(M)$, $\K = L^2(M,S) \ox \sC^2$, $\DD = \Dslash \ox 1 + i\gamma_M \ox \DD_F$  (see \cite{JGP2015} for the details).

Noncommutative geometries are equipped with a natural fermionic action defined as $S_F = (\psi,\DD \psi)$, for $\psi \in \K' \subseteq \K$ \cite{ConnesAction,ConnesLott,Barrett,KoenAction}. If the finite spectral triple is even, which is the case here, one encounters a `fermion doubling problem' \cite{LizziFermionDoubling}. A consistent prescription to avoid the overcounting of fermionic degrees of freedom has been worked out in \cite{Barrett} and consists in projecting the elements of $\K$ onto the physical subspace $\K'$. We have $\K' = P_+ \K$, with $P_+ = \tfrac{1}{2} (1 + \gamma)$, $\gamma = \gamma_M \ox \left(\begin{smallmatrix}1 & 0\\ 0 & -1\end{smallmatrix}\right)$.  A vector in $\K'$ can thus be written as
\begin{align}\label{vectors}
\K' \ni \psi = \psi_+ \ox \left( \begin{smallmatrix} 1 \\ 0 \end{smallmatrix} \right) + \psi_- \ox \left( \begin{smallmatrix} 0 \\ 1 \end{smallmatrix} \right),
\end{align}
with $\psi_{\pm}$ denoting the chirality eigenstates. The fermionic action of the model therefore reads
\begin{align}\label{action}
S_F & = (\psi_+, \Dslash \psi_+) + (\psi_-, \Dslash \psi_-) - i \mu (\psi_+,\psi_-) + i \bar{\mu} (\psi_-,\psi_+) \notag \\
& = \int_M \Big[ \bar{\psi_-} \Dslash \psi_- + \bar{\psi_+} \Dslash \psi_+ + m (\bar{\psi_-} \psi_+ + \bar{\psi_+} \psi_-) \Big],
\end{align}
with the choice $\mu = i m \in i \sR^+$. This is indeed the action describing a single Dirac fermion of mass $m$ propagating in a curved spacetime $M$.

With $\A = \A_M \ox \A_F$ we have $P(\A) \simeq M \sqcup M \cv M \x \{-,+\}$ and the `inner' space of the model consists of just two points. We stress that, despite the commutativity of the algebra $\A$, the resulting geometry is noncommutative because of the off-diagonal operator $\DD_F$.
Since the two pure states of $\A_F$ are precisely the vector states associated with $\left( \begin{smallmatrix} 1 \\ 0 \end{smallmatrix} \right), \left( \begin{smallmatrix} 0 \\ 1 \end{smallmatrix} \right) \in \H_F$ \cite{JGP2015}, it is justified --- on the strength of formula \eqref{vectors} --- to identify the two points of the model's inner space as states of definite chirality. We thus arrive at the interpretation of the space of physical states $P(\A)$ as the space of states of a `classical' fermion --- the component $M$ describes its position in spacetime and $\F = \{-,+\}$ corresponds to its chirality.

Connes first observed \cite{ConnesZitter} that the (Euclidean version of the) above almost commutative model provides a geometric viewpoint on \textit{Zitterbewegung}. However, Connes remark was only qualitative and focused on regarding the Higgs field as a gauge boson operating on the finite space $\{-,+\}$ (see also \cite{BroutZitter}). We discovered that taking into account the Lorentzian aspects of this model reveals a deeper, quantitative, connection between geometry and the `trembling motion' of fermions (cf. \cite[Theorem 9]{JGP2015}):

\begin{thm}\label{thm:two_sheeted}
Let $\tau(\gamma)$ be the proper time along a causal curve in $M$. Two states $(p,-), (q,+)$ $\in P(\A)$ are causally related with $(p,-) \preceq (q,+)$ if and only if there exists a causal curve $\gamma$ giving $p \preceq q$ in $M$, such that $\tau(\gamma) \geq \pi/(2 \abs{\mu})$.
\end{thm}

Restoring the physical dimensions in the model (which is unambiguous as $\Dslash$ has the dimension $L^{-1}$, thus $\DD_F$ must have so), we arrive at:
\begin{align}\label{zitter}
\!\!\!\! (p,-) \preceq (q,+) && \Leftrightarrow && p \preceq q \text{ and } \tau(\gamma) \geq \pi \hbar / (2 \abs{\mu} c^2).
\end{align}

The number on the RHS of \eqref{zitter} is precisely half of the \textit{Zitterbewegung} period \eqref{zitter_period} of a Dirac fermion of mass $\abs{\mu} = m$.

It is striking to realise that the possibility of \textit{Zitterbewegung} is encoded in the geometry of a purely classical model. The bound on the frequency of the fermion's quivering is of kinematic origin -- we have invoked the action \eqref{action} only to identify the relevant degrees of freedom. The apparent abrupt change of state implied by Theorem \ref{thm:two_sheeted} becomes more transparent when one considers the subspace of mixed states of indefinite chirality. In the space $M \x [-1,+1] \subset S(\A)$ the boundary of the causal cone becomes a continuous surface \cite[Fig. 2]{JGP2015} permitting a smooth evolution of the expectation value of chirality.


The advantage of the presented model is its general covariance, which guarantees that Theorem \ref{thm:two_sheeted} applies in any globally hyperbolic spacetime $M$. But the framework of noncommutative geometry is even more flexible and allows one to accommodate, via the `fluctuations' of the Dirac operator, other fields interacting with the fermion \cite{WalterBook}.

Let $\A$ and $\K$ be as previously in this section and let $\DA = (\Dslash + \Aslash) \ox 1 + i \gamma_M \ox \left(\begin{smallmatrix}0 & \Phi\\ \bar{\Phi} & 0\end{smallmatrix}\right)$, where $\Aslash = \gamma^{\mu} A_{\mu}$ with $A_{\mu} = A_{\mu}^* \in \A_M$ and $\Phi \in \A_M$, then $(\A,\K,\DA)$ is still a Lorentzian spectral triple \cite{JGP2015}. The fermionic action \eqref{action} now reads
\begin{align}
S_F(\DA) & = \int_M \Big[ \bar{\psi_-} (\Dslash + \Aslash) \psi_- + \bar{\psi_+} (\Dslash + \Aslash) \psi_+ + \bar{\psi_-} (-i \Phi) \psi_+ + \bar{\psi_+} (i \bar{\Phi}) \psi_-) \Big].
\end{align}
We thus see that $A_\mu$ is a vector field on $M$ --- for instance the electromagnetic one --- minimally coupled to the fermion, whereas $(-i \Phi)$ is a complex scalar field interacting via a Yukawa coupling \cite{WalterBook}.

The space of pure states of the interacting model is still $M \x \{-,+\}$ as the algebra remains unaltered. On the other hand, the causal cone (and \textit{a fortiori} the causal structure) does change when $\DD$ is modified. The analogue of Theorem \ref{thm:two_sheeted} reads \cite[Theorem 16]{JGP2015}:

\begin{thm}\label{thm:zitter_gauge}
Two pure states $(p,-), (q,+) \in P(\A)$ are causally related with $(p,-) \preceq (q,+)$ if and only if there exists a causal curve $\gamma$ giving $p \preceq q$ in $M$ and such that
\begin{align}\label{zitter_g}
\int_{0}^{1} ds \abs{\Phi(\gamma(s))} \sqrt{-g_{\mu\nu} \dot{\gamma}^{\mu}(s) \dot{\gamma}^{\nu}(s)} \geq \pi/2.
\end{align}
\end{thm}

An immediate consequence of Theorem \ref{thm:zitter_gauge} is that there is no impact of the vector field on the causal relations in the almost commutative spacetime at hand. In particular, it implies that the upper bound on \textit{Zitterbewegung} frequency is not altered by the presence of an electromagnetic field. On the other hand, the scalar field $\Phi$ affects causality in a more complicated way -- the LHS  of inequality \eqref{zitter_g} can be seen as a \emph{weighted proper time}. Indeed, if $\Phi$ is constant and equal to $\mu$, formula \eqref{zitter_g} reduces to \eqref{zitter}. Such a field $\Phi$ could for example be related to the variation of the mass of a point-like particle in the Einstein frame of a tensor-scalar theory. It is also amusing to observe that the impact of $\Phi$ on the causal structure is equivalent to a conformal rescaling of the metric on $M$ by $\abs{\Phi}^{-1}$. We note that the connection of the Higgs field with conformal transformations of the space(-time) in the context of noncommutative geometry was discussed in \cite{WalterBook}, but at the level of the action.

For an actual, quantum, fermion neither the concept of localisation nor that of the proper time are well defined \cite{QIandGR}, hence formulae (\ref{zitter},\ref{zitter_g}) cannot be applied directly. Nevertheless, one can elicit some phenomenological consequences from the presented model by exploiting the fact that the causal order extends to the full space $S(\A)$.

\section*{\label{sec:experiment}Simulating almost commutative spacetimes}
In an (analogue) quantum simulation some aspects of the dynamics of a complicated quantum system are mimicked in a simpler one, which is under control \cite{QuantumSimulation}. Given an almost commutative spectral triple $(\A,\K,\DD)$, together with the fermionic action \eqref{action}, one can extend this concept to probe almost commutative spacetimes in table-top experiments. The dynamics of vectors in the physical space $\K' \subseteq \K$ governed by action \eqref{action} always takes the form of a Dirac equation with some external fields. The latter can be rewritten as a Schr\"odinger equation once a suitable frame has been chosen. If one succeeds in finding a quantum system with the analogous dynamic, then one disposes of an invertible map $f: \K \to \Hs$, which determines the correspondence between the observables and the states of the system \cite{QuantumSimul1999}. Concretely, any state of the simulator system $\varphi(t) \in \Hs$ at time instant $t$ in the laboratory frame defines a state on the algebra of the emulated almost commutative spectral triple $\rho_{\varphi(t)} \in S(\A)$,
\begin{align*}
\rho_{\varphi(t)} (a) = \int_{\Sigma_t} f(\varphi(t,x))^{\dagger} a(t,x) f(\varphi(t,x)) \, dS_t(x).
\end{align*}
$\Sigma_t$ is the $t$-slice, endowed with a measure $S_t$, determined by the chosen laboratory frame and $x$ refers to the continuous degrees of freedom of the simulator system, corresponding to the space variable on $\Sigma_t \subseteq M$.

We claim that, within the domain of applicability of the simulation, the intrinsic geometry of the almost commutative spacetime should manifest itself in the simulator system. In particular, we expect the evolution of states to be causal in the sense of Definition \ref{def:causcone}, i.e.
\begin{align}\label{mixed}
\rho_{\varphi(s)} \preceq \rho_{\varphi(t)},
\end{align}
for $s \leq t$ and any initial state $\varphi(0) \in \Hs$. The details on the application of the abstract Definition \ref{def:causcone} in the wave packet formalism are explained in \cite{2NEW2016}.

The quantum simulation of a single free Dirac fermion in a flat 2-dimensional spacetime has been successfully accomplished with cold atoms \cite{ZitterBEC}, trapped ions \cite{ZitterNature} and photonic systems \cite{ZitterPhotonics}. Furthermore, a suitable experimental setup has been proposed using superconductors \cite{DiracSimulSupercond}, semiconductors \cite{DiracSimulSemicond,DiracSimulSemicond2} and graphene \cite{DiracSimulGraphene}. In the trapped-ion setting \cite{ZitterNature,QuantumSimulDirac}, the mass of the simulated fermion can be introduced dynamically, what enables a simulation of a Dirac fermion coupled to a (real) scalar field. Moreover, a possibility of studying the impact of an external electromagnetic field on \textit{Zitterbewegung} using the framework of \cite{ZitterNature} was suggested in \cite{ZitterMagnetic}.

In the presented almost commutative model, the general formula \eqref{mixed} is less explicit then the ones derived for the pure states (\ref{zitter},\ref{zitter_g}). It reflects the fact that \textit{Zitterbewegung} in the wave packet formalism is not a single-frequency oscillation \cite{ZitterPark}. Nevertheless, given concrete initial and final states of a simulator system one can unravel the consequences of formula \eqref{mixed} drawing from the fact that the causal cone is completely characterised in \cite{JGP2015} and suitable computational tools to handle mixed states were devised in \cite{EcksteinMiller2015}. One immediate upshot is that the causal relation \eqref{mixed} does not depend on the electromagnetic field (cf. Theorem \ref{thm:zitter_gauge}). Note also that the lack of \textit{Zitterbewegung} for purely positive-energy states is consistent with \eqref{mixed}. This follows directly from Theorem \ref{thm:noviolation}, the proof of which \cite{PROC2015} can be directly extended to encompass the mixed states.

\section*{\label{sec:outlook}Outlook}
We have shown that the possibility of \textit{Zitterbewegung} is encoded in the geometry of an almost commutative spacetime of a `classical' massive fermion. The presence of the electromagnetic and a Higgs-like field affects the geometry, with the latter modifying the bound on the frequency of the fermion's quivering. We argued that the consequences of this model can be tested in a suitable quantum simulation.

For a free electron the period of \textit{Zitterbewegung} \eqref{zitter_period} is of the order of $10^{-20}$ s, which is well beyond the currently available experimental time resolution. Moreover, to model a genuine electron one would need to employ the quantum field theoretic description, which seems to exclude \textit{Zitterbewegung} \cite{ZitterQFT}, at least in flat spacetimes \cite{ZitterCosmo}. The scheme presented in this Letter suggests, however, that the very foundations of quantum gauge theories might need to be refined. The principle of micro-causality --- requiring the observables in space-like separated regions to commute --- is at the core of all axiomatic approaches to quantum field theory \cite{Haag,Wightman}. If the fields have additional degrees of freedom then their background geometry is that of an almost commutative spacetime. Consequently, when constructing a quantum theory of fields, one should take into account its inherent causal structure.

The concept of causality in the space of states is at the core of the presented model. When applied to other almost commutative spacetimes (see \cite{SIGMA2014} for another gauge model), it might cast a new light on such perplexing phenomena as neutrino or quark mixing, which also involve a `motion' in the fermion's internal space. Finally, one can reach beyond the almost commutative setting and study the causal structure of genuinely noncommutative spacetimes \cite{causMoyal}. This perspective suggests that, contrary to the common belief \cite{DSR,QuantumCausality}, causal structure need not breakdown at the Planck scale, but the very notion of spacetime geometry needs to be refined.


\begin{acknowledgments}
We are indebted to Koen van den Dungen, Christoph Stephan and Walter van Suijlekom for valuable discussions. We also thank Andr\'{e} F\"{u}zfa and Henryk Arod\'{z} for comments on the manuscript. ME and NF would like to thank the Hausdorff Institute for hospitality during the Trimester Program ``Non-commutative Geometry and its Applications''. ME was supported by the Foundation for Polish Science under the programme START 2016. ME acknowledges the support of the Marian Smoluchowski Krak\'{o}w Research Consortium ``Matter--Energy--Future'' within the programme KNOW.
\end{acknowledgments}

\bibliography{causality_bib}{}

\begin{thebibliography}{64}%
\makeatletter
\providecommand \@ifxundefined [1]{%
 \@ifx{#1\undefined}
}%
\providecommand \@ifnum [1]{%
 \ifnum #1\expandafter \@firstoftwo
 \else \expandafter \@secondoftwo
 \fi
}%
\providecommand \@ifx [1]{%
 \ifx #1\expandafter \@firstoftwo
 \else \expandafter \@secondoftwo
 \fi
}%
\providecommand \natexlab [1]{#1}%
\providecommand \enquote  [1]{``#1''}%
\providecommand \bibnamefont  [1]{#1}%
\providecommand \bibfnamefont [1]{#1}%
\providecommand \citenamefont [1]{#1}%
\providecommand \href@noop [0]{\@secondoftwo}%
\providecommand \href [0]{\begingroup \@sanitize@url \@href}%
\providecommand \@href[1]{\@@startlink{#1}\@@href}%
\providecommand \@@href[1]{\endgroup#1\@@endlink}%
\providecommand \@sanitize@url [0]{\catcode `\\12\catcode `\$12\catcode
  `\&12\catcode `\#12\catcode `\^12\catcode `\_12\catcode `\%12\relax}%
\providecommand \@@startlink[1]{}%
\providecommand \@@endlink[0]{}%
\providecommand \url  [0]{\begingroup\@sanitize@url \@url }%
\providecommand \@url [1]{\endgroup\@href {#1}{\urlprefix }}%
\providecommand \urlprefix  [0]{URL }%
\providecommand \Eprint [0]{\href }%
\providecommand \doibase [0]{http://dx.doi.org/}%
\providecommand \selectlanguage [0]{\@gobble}%
\providecommand \bibinfo  [0]{\@secondoftwo}%
\providecommand \bibfield  [0]{\@secondoftwo}%
\providecommand \translation [1]{[#1]}%
\providecommand \BibitemOpen [0]{}%
\providecommand \bibitemStop [0]{}%
\providecommand \bibitemNoStop [0]{.\EOS\space}%
\providecommand \EOS [0]{\spacefactor3000\relax}%
\providecommand \BibitemShut  [1]{\csname bibitem#1\endcsname}%
\let\auto@bib@innerbib\@empty
\bibitem [{\citenamefont {Dirac}(1958)}]{DiracBook}%
  \BibitemOpen
  \bibfield  {author} {\bibinfo {author} {\bibfnamefont {P.~A.~M.}\
  \bibnamefont {Dirac}},\ }\href@noop {} {\emph {\bibinfo {title} {The
  principles of quantum mechanics}}},\ Vol.~\bibinfo {volume} {4}\ (\bibinfo
  {publisher} {Clarendon Press Oxford},\ \bibinfo {year} {1958})\BibitemShut
  {NoStop}%
\bibitem [{\citenamefont {Friedrich}\ and\ \citenamefont
  {Nestke}(2000)}]{Friedrich}%
  \BibitemOpen
  \bibfield  {author} {\bibinfo {author} {\bibfnamefont {T.}~\bibnamefont
  {Friedrich}}\ and\ \bibinfo {author} {\bibfnamefont {A.}~\bibnamefont
  {Nestke}},\ }\href@noop {} {\emph {\bibinfo {title} {Dirac operators in
  Riemannian geometry}}},\ \bibinfo {series} {Graduate Studies in Mathematics},
  Vol.~\bibinfo {volume} {25}\ (\bibinfo  {publisher} {American Mathematical
  Society Providence},\ \bibinfo {year} {2000})\BibitemShut {NoStop}%
\bibitem [{\citenamefont {Connes}(1994)}]{C94}%
  \BibitemOpen
  \bibfield  {author} {\bibinfo {author} {\bibfnamefont {A.}~\bibnamefont
  {Connes}},\ }\href@noop {} {\emph {\bibinfo {title} {Noncommutative
  Geometry}}}\ (\bibinfo  {publisher} {Academic Press},\ \bibinfo {year}
  {1994})\BibitemShut {NoStop}%
\bibitem [{\citenamefont {Connes}\ and\ \citenamefont {Marcolli}(2008)}]{MC08}%
  \BibitemOpen
  \bibfield  {author} {\bibinfo {author} {\bibfnamefont {A.}~\bibnamefont
  {Connes}}\ and\ \bibinfo {author} {\bibfnamefont {M.}~\bibnamefont
  {Marcolli}},\ }\href@noop {} {\emph {\bibinfo {title} {Noncommutative
  Geometry, Quantum Fields and Motives}}},\ \bibinfo {series} {Colloquium
  Publications}, Vol.~\bibinfo {volume} {55}\ (\bibinfo  {publisher} {American
  Mathematical Society},\ \bibinfo {year} {2008})\BibitemShut {NoStop}%
\bibitem [{\citenamefont {Franco}\ and\ \citenamefont
  {Eckstein}(2013)}]{CQG2013}%
  \BibitemOpen
  \bibfield  {author} {\bibinfo {author} {\bibfnamefont {N.}~\bibnamefont
  {Franco}}\ and\ \bibinfo {author} {\bibfnamefont {M.}~\bibnamefont
  {Eckstein}},\ }\href {\doibase 10.1088/0264-9381/30/13/135007} {\bibfield
  {journal} {\bibinfo  {journal} {Classical and Quantum Gravity}\ }\textbf
  {\bibinfo {volume} {30}},\ \bibinfo {pages} {135007} (\bibinfo {year}
  {2013})},\ \Eprint {http://arxiv.org/abs/arXiv:1212.5171 [math-ph]}
  {arXiv:1212.5171 [math-ph]} \BibitemShut {NoStop}%
\bibitem [{\citenamefont {Chamseddine}\ and\ \citenamefont
  {Connes}(1996)}]{CC96}%
  \BibitemOpen
  \bibfield  {author} {\bibinfo {author} {\bibfnamefont {A.~H.}\ \bibnamefont
  {Chamseddine}}\ and\ \bibinfo {author} {\bibfnamefont {A.}~\bibnamefont
  {Connes}},\ }\href {\doibase 10.1103/PhysRevLett.77.4868} {\bibfield
  {journal} {\bibinfo  {journal} {Physical Review Letters}\ }\textbf {\bibinfo
  {volume} {77}},\ \bibinfo {pages} {4868} (\bibinfo {year} {1996})},\ \Eprint
  {http://arxiv.org/abs/arXiv:hep-th/9606056} {arXiv:hep-th/9606056}
  \BibitemShut {NoStop}%
\bibitem [{\citenamefont {Chamseddine}\ \emph {et~al.}(2007)\citenamefont
  {Chamseddine}, \citenamefont {Connes},\ and\ \citenamefont
  {Marcolli}}]{Almost1}%
  \BibitemOpen
  \bibfield  {author} {\bibinfo {author} {\bibfnamefont {A.~H.}\ \bibnamefont
  {Chamseddine}}, \bibinfo {author} {\bibfnamefont {A.}~\bibnamefont {Connes}},
  \ and\ \bibinfo {author} {\bibfnamefont {M.}~\bibnamefont {Marcolli}},\
  }\href {http://projecteuclid.org/euclid.atmp/1198095373} {\bibfield
  {journal} {\bibinfo  {journal} {Advances in Theoretical and Mathematical
  Physics}\ }\textbf {\bibinfo {volume} {11}},\ \bibinfo {pages} {991}
  (\bibinfo {year} {2007})},\ \Eprint
  {http://arxiv.org/abs/arXiv:hep-th/0610241} {arXiv:hep-th/0610241}
  \BibitemShut {NoStop}%
\bibitem [{\citenamefont {van Suijlekom}(2015)}]{WalterBook}%
  \BibitemOpen
  \bibfield  {author} {\bibinfo {author} {\bibfnamefont {W.~D.}\ \bibnamefont
  {van Suijlekom}},\ }\href {\doibase 10.1007/978-94-017-9162-5} {\emph
  {\bibinfo {title} {Noncommutative Geometry and Particle Physics}}},\
  Mathematical Physics Studies\ (\bibinfo  {publisher} {Springer},\ \bibinfo
  {year} {2015})\BibitemShut {NoStop}%
\bibitem [{\citenamefont {Marcolli}(2011)}]{MarcolliCosmSurv}%
  \BibitemOpen
  \bibfield  {author} {\bibinfo {author} {\bibfnamefont {M.}~\bibnamefont
  {Marcolli}},\ }\href {\doibase 10.1142/S0219887811005592} {\bibfield
  {journal} {\bibinfo  {journal} {International Journal of Geometric Methods in
  Modern Physics}\ }\textbf {\bibinfo {volume} {08}},\ \bibinfo {pages} {1131}
  (\bibinfo {year} {2011})}\BibitemShut {NoStop}%
\bibitem [{\citenamefont {Lambiase}\ \emph {et~al.}(2013)\citenamefont
  {Lambiase}, \citenamefont {Sakellariadou},\ and\ \citenamefont
  {Stabile}}]{MairiJCAP}%
  \BibitemOpen
  \bibfield  {author} {\bibinfo {author} {\bibfnamefont {G.}~\bibnamefont
  {Lambiase}}, \bibinfo {author} {\bibfnamefont {M.}~\bibnamefont
  {Sakellariadou}}, \ and\ \bibinfo {author} {\bibfnamefont {A.}~\bibnamefont
  {Stabile}},\ }\href {\doibase 10.1088/1475-7516/2013/12/020} {\bibfield
  {journal} {\bibinfo  {journal} {Journal of Cosmology and Astroparticle
  Physics}\ }\textbf {\bibinfo {volume} {2013}},\ \bibinfo {pages} {020}
  (\bibinfo {year} {2013})},\ \Eprint {http://arxiv.org/abs/arXiv:1302.2336
  [gr-qc]} {arXiv:1302.2336 [gr-qc]} \BibitemShut {NoStop}%
\bibitem [{\citenamefont {Connes}(1988)}]{ConnesZitter}%
  \BibitemOpen
  \bibfield  {author} {\bibinfo {author} {\bibfnamefont {A.}~\bibnamefont
  {Connes}},\ }in\ \href {\doibase 10.1007/978-1-4613-0729-7_3} {\emph
  {\bibinfo {booktitle} {Nonperturbative Quantum Field Teory}}},\ \bibinfo
  {series} {NATO ASI Series B: Physics}, Vol.\ \bibinfo {volume} {185},\
  \bibinfo {editor} {edited by\ \bibinfo {editor} {\bibfnamefont
  {G.}~\bibnamefont {'t~Hooft}}, \bibinfo {editor} {\bibfnamefont
  {A.}~\bibnamefont {Jaffe}}, \bibinfo {editor} {\bibfnamefont
  {G.}~\bibnamefont {Mack}}, \bibinfo {editor} {\bibfnamefont {P.}~\bibnamefont
  {Mitter}}, \ and\ \bibinfo {editor} {\bibfnamefont {R.}~\bibnamefont
  {Stora}}}\ (\bibinfo  {publisher} {Plenum Press, New York},\ \bibinfo {year}
  {1988})\ pp.\ \bibinfo {pages} {33--70},\ \bibinfo {note} {proceedings of a
  NATO Advanced Study Institute on Nonperturbative Quantum Field Theory, held
  July 16-30, 1987, in Carg\`ese, France}\BibitemShut {NoStop}%
\bibitem [{\citenamefont {Schr\"odinger}(1930)}]{ZitterSchroedinger}%
  \BibitemOpen
  \bibfield  {author} {\bibinfo {author} {\bibfnamefont {E.}~\bibnamefont
  {Schr\"odinger}},\ }\href@noop {} {\bibfield  {journal} {\bibinfo  {journal}
  {Sitzungsberichte der Preu$\mathrm{beta}$ischen Akademie der Wissenschaften.
  Physikalisch-mathematische Klasse}\ }\textbf {\bibinfo {volume} {24}},\
  \bibinfo {pages} {418} (\bibinfo {year} {1930})}\BibitemShut {NoStop}%
\bibitem [{\citenamefont {Georgescu}\ \emph {et~al.}(2014)\citenamefont
  {Georgescu}, \citenamefont {Ashhab},\ and\ \citenamefont
  {Nori}}]{QuantumSimulation}%
  \BibitemOpen
  \bibfield  {author} {\bibinfo {author} {\bibfnamefont {I.}~\bibnamefont
  {Georgescu}}, \bibinfo {author} {\bibfnamefont {S.}~\bibnamefont {Ashhab}}, \
  and\ \bibinfo {author} {\bibfnamefont {F.}~\bibnamefont {Nori}},\ }\href
  {\doibase 10.1103/RevModPhys.86.153} {\bibfield  {journal} {\bibinfo
  {journal} {Reviews of Modern Physics}\ }\textbf {\bibinfo {volume} {86}},\
  \bibinfo {pages} {153} (\bibinfo {year} {2014})},\ \Eprint
  {http://arxiv.org/abs/arXiv:1308.6253 [quant-ph]} {arXiv:1308.6253
  [quant-ph]} \BibitemShut {NoStop}%
\bibitem [{\citenamefont {Thaller}(1992)}]{Thaller}%
  \BibitemOpen
  \bibfield  {author} {\bibinfo {author} {\bibfnamefont {B.}~\bibnamefont
  {Thaller}},\ }\href {\doibase 10.1007/978-3-662-02753-0} {\emph {\bibinfo
  {title} {The {Dirac} {E}quation}}},\ \bibinfo {series} {Theoretical and
  Mathematical Physics}, Vol.~\bibinfo {volume} {31}\ (\bibinfo  {publisher}
  {Springer-Verlag Berlin},\ \bibinfo {year} {1992})\BibitemShut {NoStop}%
\bibitem [{\citenamefont {LeBlanc}\ \emph {et~al.}(2013)\citenamefont
  {LeBlanc}, \citenamefont {Beeler}, \citenamefont {Jim{\'e}nez-Garc{\'\i}a},
  \citenamefont {Perry}, \citenamefont {Sugawa}, \citenamefont {Williams},\
  and\ \citenamefont {Spielman}}]{ZitterBEC}%
  \BibitemOpen
  \bibfield  {author} {\bibinfo {author} {\bibfnamefont {L.}~\bibnamefont
  {LeBlanc}}, \bibinfo {author} {\bibfnamefont {M.}~\bibnamefont {Beeler}},
  \bibinfo {author} {\bibfnamefont {K.}~\bibnamefont
  {Jim{\'e}nez-Garc{\'\i}a}}, \bibinfo {author} {\bibfnamefont
  {A.}~\bibnamefont {Perry}}, \bibinfo {author} {\bibfnamefont
  {S.}~\bibnamefont {Sugawa}}, \bibinfo {author} {\bibfnamefont
  {R.}~\bibnamefont {Williams}}, \ and\ \bibinfo {author} {\bibfnamefont
  {I.}~\bibnamefont {Spielman}},\ }\href {\doibase
  10.1088/1367-2630/15/7/073011} {\bibfield  {journal} {\bibinfo  {journal}
  {New Journal of Physics}\ }\textbf {\bibinfo {volume} {15}},\ \bibinfo
  {pages} {073011} (\bibinfo {year} {2013})},\ \Eprint
  {http://arxiv.org/abs/arXiv:1303.0914 [cond-mat.quant-gas]} {arXiv:1303.0914
  [cond-mat.quant-gas]} \BibitemShut {NoStop}%
\bibitem [{\citenamefont {Park}(2012)}]{ZitterPark}%
  \BibitemOpen
  \bibfield  {author} {\bibinfo {author} {\bibfnamefont {S.~T.}\ \bibnamefont
  {Park}},\ }\href {\doibase 10.1103/PhysRevA.86.062105} {\bibfield  {journal}
  {\bibinfo  {journal} {Physical Review A}\ }\textbf {\bibinfo {volume} {86}},\
  \bibinfo {pages} {062105} (\bibinfo {year} {2012})}\BibitemShut {NoStop}%
\bibitem [{\citenamefont {Huang}(1952)}]{ZitterHuang}%
  \BibitemOpen
  \bibfield  {author} {\bibinfo {author} {\bibfnamefont {K.}~\bibnamefont
  {Huang}},\ }\href {\doibase 10.1119/1.1933296} {\bibfield  {journal}
  {\bibinfo  {journal} {American Journal of Physics}\ }\textbf {\bibinfo
  {volume} {20}},\ \bibinfo {pages} {479} (\bibinfo {year} {1952})}\BibitemShut
  {NoStop}%
\bibitem [{\citenamefont {Krekora}\ \emph {et~al.}(2004)\citenamefont
  {Krekora}, \citenamefont {Su},\ and\ \citenamefont {Grobe}}]{ZitterQFT}%
  \BibitemOpen
  \bibfield  {author} {\bibinfo {author} {\bibfnamefont {P.}~\bibnamefont
  {Krekora}}, \bibinfo {author} {\bibfnamefont {Q.}~\bibnamefont {Su}}, \ and\
  \bibinfo {author} {\bibfnamefont {R.}~\bibnamefont {Grobe}},\ }\href
  {\doibase http://dx.doi.org/10.1103/PhysRevLett.93.043004} {\bibfield
  {journal} {\bibinfo  {journal} {Physical Review Letters}\ }\textbf {\bibinfo
  {volume} {93}},\ \bibinfo {pages} {043004} (\bibinfo {year}
  {2004})}\BibitemShut {NoStop}%
\bibitem [{\citenamefont {Kobakhidze}\ \emph {et~al.}(2016)\citenamefont
  {Kobakhidze}, \citenamefont {Manning},\ and\ \citenamefont
  {Tureanu}}]{ZitterCosmo}%
  \BibitemOpen
  \bibfield  {author} {\bibinfo {author} {\bibfnamefont {A.}~\bibnamefont
  {Kobakhidze}}, \bibinfo {author} {\bibfnamefont {A.}~\bibnamefont {Manning}},
  \ and\ \bibinfo {author} {\bibfnamefont {A.}~\bibnamefont {Tureanu}},\ }\href
  {\doibase http://dx.doi.org/10.1016/j.physletb.2016.03.049} {\bibfield
  {journal} {\bibinfo  {journal} {Physics Letters B}\ }\textbf {\bibinfo
  {volume} {757}},\ \bibinfo {pages} {84 } (\bibinfo {year} {2016})},\ \Eprint
  {http://arxiv.org/abs/arXiv:1508.06322 [hep-th]} {arXiv:1508.06322 [hep-th]}
  \BibitemShut {NoStop}%
\bibitem [{\citenamefont {Gerritsma}\ \emph {et~al.}(2010)\citenamefont
  {Gerritsma}, \citenamefont {Kirchmair}, \citenamefont {Zaehringer},
  \citenamefont {Solano}, \citenamefont {Blatt},\ and\ \citenamefont
  {Roos}}]{ZitterNature}%
  \BibitemOpen
  \bibfield  {author} {\bibinfo {author} {\bibfnamefont {R.}~\bibnamefont
  {Gerritsma}}, \bibinfo {author} {\bibfnamefont {G.}~\bibnamefont
  {Kirchmair}}, \bibinfo {author} {\bibfnamefont {F.}~\bibnamefont
  {Zaehringer}}, \bibinfo {author} {\bibfnamefont {E.}~\bibnamefont {Solano}},
  \bibinfo {author} {\bibfnamefont {R.}~\bibnamefont {Blatt}}, \ and\ \bibinfo
  {author} {\bibfnamefont {C.}~\bibnamefont {Roos}},\ }\href {\doibase
  10.1038/nature08688} {\bibfield  {journal} {\bibinfo  {journal} {Nature}\
  }\textbf {\bibinfo {volume} {463}},\ \bibinfo {pages} {68} (\bibinfo {year}
  {2010})},\ \Eprint {http://arxiv.org/abs/arXiv:0909.0674 [quant-ph]}
  {arXiv:0909.0674 [quant-ph]} \BibitemShut {NoStop}%
\bibitem [{\citenamefont {Dreisow}\ \emph {et~al.}(2010)\citenamefont
  {Dreisow}, \citenamefont {Heinrich}, \citenamefont {Keil}, \citenamefont
  {T\"unnermann}, \citenamefont {Nolte}, \citenamefont {Longhi},\ and\
  \citenamefont {Szameit}}]{ZitterPhotonics}%
  \BibitemOpen
  \bibfield  {author} {\bibinfo {author} {\bibfnamefont {F.}~\bibnamefont
  {Dreisow}}, \bibinfo {author} {\bibfnamefont {M.}~\bibnamefont {Heinrich}},
  \bibinfo {author} {\bibfnamefont {R.}~\bibnamefont {Keil}}, \bibinfo {author}
  {\bibfnamefont {A.}~\bibnamefont {T\"unnermann}}, \bibinfo {author}
  {\bibfnamefont {S.}~\bibnamefont {Nolte}}, \bibinfo {author} {\bibfnamefont
  {S.}~\bibnamefont {Longhi}}, \ and\ \bibinfo {author} {\bibfnamefont
  {A.}~\bibnamefont {Szameit}},\ }\href {\doibase
  10.1103/PhysRevLett.105.143902} {\bibfield  {journal} {\bibinfo  {journal}
  {Physical Review Letters}\ }\textbf {\bibinfo {volume} {105}},\ \bibinfo
  {pages} {143902} (\bibinfo {year} {2010})}\BibitemShut {NoStop}%
\bibitem [{\citenamefont {Katsnelson}(2006)}]{ZitterGraphene1}%
  \BibitemOpen
  \bibfield  {author} {\bibinfo {author} {\bibfnamefont {M.~I.}\ \bibnamefont
  {Katsnelson}},\ }\href {\doibase 10.1140/epjb/e2006-00203-1} {\bibfield
  {journal} {\bibinfo  {journal} {The European Physical Journal B -- Condensed
  Matter and Complex Systems}\ }\textbf {\bibinfo {volume} {51}},\ \bibinfo
  {pages} {157} (\bibinfo {year} {2006})},\ \Eprint
  {http://arxiv.org/abs/arXiv:cond-mat/0512337 [cond-mat.mes-hall]}
  {arXiv:cond-mat/0512337 [cond-mat.mes-hall]} \BibitemShut {NoStop}%
\bibitem [{\citenamefont {Tworzyd\l{}o}\ \emph {et~al.}(2006)\citenamefont
  {Tworzyd\l{}o}, \citenamefont {Trauzettel}, \citenamefont {Titov},
  \citenamefont {Rycerz},\ and\ \citenamefont {Beenakker}}]{ZitterGraphene2}%
  \BibitemOpen
  \bibfield  {author} {\bibinfo {author} {\bibfnamefont {J.}~\bibnamefont
  {Tworzyd\l{}o}}, \bibinfo {author} {\bibfnamefont {B.}~\bibnamefont
  {Trauzettel}}, \bibinfo {author} {\bibfnamefont {M.}~\bibnamefont {Titov}},
  \bibinfo {author} {\bibfnamefont {A.}~\bibnamefont {Rycerz}}, \ and\ \bibinfo
  {author} {\bibfnamefont {C.~W.~J.}\ \bibnamefont {Beenakker}},\ }\href
  {\doibase 10.1103/PhysRevLett.96.246802} {\bibfield  {journal} {\bibinfo
  {journal} {Physical Review Letters}\ }\textbf {\bibinfo {volume} {96}},\
  \bibinfo {pages} {246802} (\bibinfo {year} {2006})},\ \Eprint
  {http://arxiv.org/abs/arXiv:cond-mat/0603315 [cond-mat.mes-hall]}
  {arXiv:cond-mat/0603315 [cond-mat.mes-hall]} \BibitemShut {NoStop}%
\bibitem [{\citenamefont {De~Leo}\ and\ \citenamefont
  {Rotelli}(1998)}]{ChiralZitter}%
  \BibitemOpen
  \bibfield  {author} {\bibinfo {author} {\bibfnamefont {S.}~\bibnamefont
  {De~Leo}}\ and\ \bibinfo {author} {\bibfnamefont {P.}~\bibnamefont
  {Rotelli}},\ }\href {\doibase 10.1023/A:1026602305499} {\bibfield  {journal}
  {\bibinfo  {journal} {International Journal of Theoretical Physics}\ }\textbf
  {\bibinfo {volume} {37}},\ \bibinfo {pages} {2193} (\bibinfo {year}
  {1998})},\ \Eprint {http://arxiv.org/abs/arXiv:hep-ph/9605255}
  {arXiv:hep-ph/9605255} \BibitemShut {NoStop}%
\bibitem [{\citenamefont {Bernardini}(2007)}]{ChiralZitter2}%
  \BibitemOpen
  \bibfield  {author} {\bibinfo {author} {\bibfnamefont {A.~E.}\ \bibnamefont
  {Bernardini}},\ }\href {\doibase 10.1140/epjc/s10052-007-0222-x} {\bibfield
  {journal} {\bibinfo  {journal} {The European Physical Journal C-Particles and
  Fields}\ }\textbf {\bibinfo {volume} {50}},\ \bibinfo {pages} {673} (\bibinfo
  {year} {2007})},\ \Eprint {http://arxiv.org/abs/arXiv:hep-th/0701091}
  {arXiv:hep-th/0701091} \BibitemShut {NoStop}%
\bibitem [{\citenamefont {Hestenes}(1990)}]{Hestenes1990}%
  \BibitemOpen
  \bibfield  {author} {\bibinfo {author} {\bibfnamefont {D.}~\bibnamefont
  {Hestenes}},\ }\href {\doibase 10.1007/BF01889466} {\bibfield  {journal}
  {\bibinfo  {journal} {Foundations of Physics}\ }\textbf {\bibinfo {volume}
  {20}},\ \bibinfo {pages} {1213} (\bibinfo {year} {1990})}\BibitemShut
  {NoStop}%
\bibitem [{\citenamefont {Hestenes}(2003)}]{Hestenes2003}%
  \BibitemOpen
  \bibfield  {author} {\bibinfo {author} {\bibfnamefont {D.}~\bibnamefont
  {Hestenes}},\ }\href {http://aflb.ensmp.fr/AFLB-283/aflb283p367.pdf}
  {\bibfield  {journal} {\bibinfo  {journal} {Annales de la Fondation Louis de
  Broglie}\ }\textbf {\bibinfo {volume} {28}},\ \bibinfo {pages} {390}
  (\bibinfo {year} {2003})}\BibitemShut {NoStop}%
\bibitem [{\citenamefont {Hestenes}(2010)}]{Hestenes2010}%
  \BibitemOpen
  \bibfield  {author} {\bibinfo {author} {\bibfnamefont {D.}~\bibnamefont
  {Hestenes}},\ }\href {\doibase 10.1007/s10701-009-9360-3} {\bibfield
  {journal} {\bibinfo  {journal} {Foundations of Physics}\ }\textbf {\bibinfo
  {volume} {40}},\ \bibinfo {pages} {1} (\bibinfo {year} {2010})}\BibitemShut
  {NoStop}%
\bibitem [{\citenamefont {Penrose}(2004)}]{PenroseBook}%
  \BibitemOpen
  \bibfield  {author} {\bibinfo {author} {\bibfnamefont {R.}~\bibnamefont
  {Penrose}},\ }\href@noop {} {\emph {\bibinfo {title} {The Road to Reality: a
  Complete Guide to the Laws of the Universe}}}\ (\bibinfo  {publisher}
  {Jonathan Cape, London},\ \bibinfo {address} {London},\ \bibinfo {year}
  {2004})\BibitemShut {NoStop}%
\bibitem [{\citenamefont {Penrose}(1997)}]{PenroseSpin}%
  \BibitemOpen
  \bibfield  {author} {\bibinfo {author} {\bibfnamefont {R.}~\bibnamefont
  {Penrose}},\ }\href {\doibase 10.1088/0143-0807/18/3/006} {\bibfield
  {journal} {\bibinfo  {journal} {European Journal of Physics}\ }\textbf
  {\bibinfo {volume} {18}},\ \bibinfo {pages} {164} (\bibinfo {year}
  {1997})}\BibitemShut {NoStop}%
\bibitem [{\citenamefont {Strohmaier}(2006)}]{Stro}%
  \BibitemOpen
  \bibfield  {author} {\bibinfo {author} {\bibfnamefont {A.}~\bibnamefont
  {Strohmaier}},\ }\href {\doibase 10.1016/j.geomphys.2005.01.005} {\bibfield
  {journal} {\bibinfo  {journal} {Journal of Geometry and Physics}\ }\textbf
  {\bibinfo {volume} {56}},\ \bibinfo {pages} {175} (\bibinfo {year} {2006})},\
  \Eprint {http://arxiv.org/abs/arXiv:math-ph/0110001} {arXiv:math-ph/0110001}
  \BibitemShut {NoStop}%
\bibitem [{\citenamefont {Paschke}\ and\ \citenamefont {Sitarz}(2006)}]{Pas}%
  \BibitemOpen
  \bibfield  {author} {\bibinfo {author} {\bibfnamefont {M.}~\bibnamefont
  {Paschke}}\ and\ \bibinfo {author} {\bibfnamefont {A.}~\bibnamefont
  {Sitarz}},\ }\href@noop {} {\enquote {\bibinfo {title} {Equivariant
  {L}orentzian spectral triples},}\ } (\bibinfo {year} {2006}),\ \bibinfo
  {note} {preprint
  \href{https://arxiv.org/abs/math-ph/0611029}{arXiv:math-ph/0611029}}\BibitemShut
  {NoStop}%
\bibitem [{\citenamefont {Franco}(2011)}]{F4}%
  \BibitemOpen
  \bibfield  {author} {\bibinfo {author} {\bibfnamefont {N.}~\bibnamefont
  {Franco}},\ }\emph {\bibinfo {title} {Lorentzian Approach to Noncommutative
  Geometry}},\ \href@noop {} {Ph.D. thesis},\ \bibinfo  {school} {University of
  Namur FUNDP} (\bibinfo {year} {2011}),\ \Eprint
  {http://arxiv.org/abs/arXiv:1108.0592 [math-ph]} {arXiv:1108.0592 [math-ph]}
  \BibitemShut {NoStop}%
\bibitem [{\citenamefont {Verch}(2011)}]{Verch11}%
  \BibitemOpen
  \bibfield  {author} {\bibinfo {author} {\bibfnamefont {R.}~\bibnamefont
  {Verch}},\ }in\ \href {\doibase 10.5506/APhysPolBSupp.4.507} {\emph {\bibinfo
  {booktitle} {Acta Physica Polonica B Proceedings Supplement}}},\
  Vol.~\bibinfo {volume} {4}\ (\bibinfo {year} {2011})\ pp.\ \bibinfo {pages}
  {507--530},\ \Eprint {http://arxiv.org/abs/arXiv:1106.1138 [math-ph]}
  {arXiv:1106.1138 [math-ph]} \BibitemShut {NoStop}%
\bibitem [{\citenamefont {van~den Dungen}\ \emph {et~al.}(2013)\citenamefont
  {van~den Dungen}, \citenamefont {Paschke},\ and\ \citenamefont
  {Rennie}}]{Rennie12}%
  \BibitemOpen
  \bibfield  {author} {\bibinfo {author} {\bibfnamefont {K.}~\bibnamefont
  {van~den Dungen}}, \bibinfo {author} {\bibfnamefont {M.}~\bibnamefont
  {Paschke}}, \ and\ \bibinfo {author} {\bibfnamefont {A.}~\bibnamefont
  {Rennie}},\ }\href {\doibase 10.1016/j.geomphys.2013.04.011} {\bibfield
  {journal} {\bibinfo  {journal} {Journal of Geometry and Physics}\ }\textbf
  {\bibinfo {volume} {73}},\ \bibinfo {pages} {37} (\bibinfo {year} {2013})},\
  \Eprint {http://arxiv.org/abs/arXiv:1207.2112 [math.OA]} {arXiv:1207.2112
  [math.OA]} \BibitemShut {NoStop}%
\bibitem [{\citenamefont {Franco}(2014)}]{F5}%
  \BibitemOpen
  \bibfield  {author} {\bibinfo {author} {\bibfnamefont {N.}~\bibnamefont
  {Franco}},\ }\href {\doibase 10.1142/S0129055X14300076} {\bibfield  {journal}
  {\bibinfo  {journal} {Reviews in Mathematical Physics}\ }\textbf {\bibinfo
  {volume} {26}},\ \bibinfo {pages} {1430007} (\bibinfo {year} {2014})},\
  \Eprint {http://arxiv.org/abs/arXiv:1210.6575 [math-ph]} {arXiv:1210.6575
  [math-ph]} \BibitemShut {NoStop}%
\bibitem [{\citenamefont {Bogn{\'{a}}r}(1974)}]{Bog}%
  \BibitemOpen
  \bibfield  {author} {\bibinfo {author} {\bibfnamefont {J.}~\bibnamefont
  {Bogn{\'{a}}r}},\ }\href {\doibase 10.1007/978-3-642-65567-8} {\emph
  {\bibinfo {title} {{Indefinite Inner Product Spaces}}}}\ (\bibinfo
  {publisher} {Springer},\ \bibinfo {year} {1974})\BibitemShut {NoStop}%
\bibitem [{\citenamefont {Franco}\ and\ \citenamefont
  {Eckstein}(2014)}]{SIGMA2014}%
  \BibitemOpen
  \bibfield  {author} {\bibinfo {author} {\bibfnamefont {N.}~\bibnamefont
  {Franco}}\ and\ \bibinfo {author} {\bibfnamefont {M.}~\bibnamefont
  {Eckstein}},\ }\href {\doibase 10.3842/SIGMA.2014.010} {\bibfield  {journal}
  {\bibinfo  {journal} {Symmetry, Integrability and Geometry: Methods and
  Applications}\ }\textbf {\bibinfo {volume} {10}},\ \bibinfo {pages} {010}
  (\bibinfo {year} {2014})},\ \bibinfo {note} {special Issue on Noncommutative
  Geometry and Quantum Groups in honor of Marc A. Rieffel.},\ \Eprint
  {http://arxiv.org/abs/arXiv:1310.8225 [math-ph]} {arXiv:1310.8225 [math-ph]}
  \BibitemShut {NoStop}%
\bibitem [{\citenamefont {Haag}(1996)}]{Haag}%
  \BibitemOpen
  \bibfield  {author} {\bibinfo {author} {\bibfnamefont {R.}~\bibnamefont
  {Haag}},\ }\href {https://books.google.pl/books?id=VdyJQgAACAAJ} {\emph
  {\bibinfo {title} {Local {Q}uantum {P}hysics: {F}ields, {P}articles,
  {A}lgebras}}},\ Theoretical and Mathematical Physics\ (\bibinfo  {publisher}
  {Springer Berlin Heidelberg},\ \bibinfo {year} {1996})\BibitemShut {NoStop}%
\bibitem [{\citenamefont {Strocchi}(2008)}]{Strocchi}%
  \BibitemOpen
  \bibfield  {author} {\bibinfo {author} {\bibfnamefont {F.}~\bibnamefont
  {Strocchi}},\ }\href {\doibase 10.1142/5908} {\emph {\bibinfo {title} {An
  Introduction to the Mathematical Structure of Quantum Mechanics}}}\ (\bibinfo
   {publisher} {World Scientific},\ \bibinfo {year} {2008})\BibitemShut
  {NoStop}%
\bibitem [{\citenamefont {Keyl}(2002)}]{Keyl}%
  \BibitemOpen
  \bibfield  {author} {\bibinfo {author} {\bibfnamefont {M.}~\bibnamefont
  {Keyl}},\ }\href {\doibase http://dx.doi.org/10.1016/S0370-1573(02)00266-1}
  {\bibfield  {journal} {\bibinfo  {journal} {Physics Reports}\ }\textbf
  {\bibinfo {volume} {369}},\ \bibinfo {pages} {431 } (\bibinfo {year}
  {2002})},\ \Eprint {http://arxiv.org/abs/arXiv:quant-ph/0202122}
  {arXiv:quant-ph/0202122} \BibitemShut {NoStop}%
\bibitem [{\citenamefont {Eckstein}\ and\ \citenamefont
  {Franco}(2015)}]{PROC2015}%
  \BibitemOpen
  \bibfield  {author} {\bibinfo {author} {\bibfnamefont {M.}~\bibnamefont
  {Eckstein}}\ and\ \bibinfo {author} {\bibfnamefont {N.}~\bibnamefont
  {Franco}},\ }in\ \href@noop {} {\emph {\bibinfo {booktitle} {Frontiers of
  Fundamental Physics 14}}}\ (\bibinfo {year} {2015})\ \bibinfo {note}
  {\href{http://pos.sissa.it/cgi-bin/reader/contribution.cgi?id=PoS(FFP14)138}{PoS(FFP14)138}}\BibitemShut
  {NoStop}%
\bibitem [{\citenamefont {Kadison}\ and\ \citenamefont
  {Ringrose}(1986)}]{Kadison}%
  \BibitemOpen
  \bibfield  {author} {\bibinfo {author} {\bibfnamefont {R.~V.}\ \bibnamefont
  {Kadison}}\ and\ \bibinfo {author} {\bibfnamefont {J.~R.}\ \bibnamefont
  {Ringrose}},\ }\href@noop {} {\emph {\bibinfo {title} {Fundamentals of the
  Theory of Operator Algebras}}},\ Vol.\ \bibinfo {volume} {2. Advanced
  Theory}\ (\bibinfo  {publisher} {Academic Press},\ \bibinfo {year}
  {1986})\BibitemShut {NoStop}%
\bibitem [{\citenamefont {Franco}\ and\ \citenamefont
  {Eckstein}(2015)}]{JGP2015}%
  \BibitemOpen
  \bibfield  {author} {\bibinfo {author} {\bibfnamefont {N.}~\bibnamefont
  {Franco}}\ and\ \bibinfo {author} {\bibfnamefont {M.}~\bibnamefont
  {Eckstein}},\ }\href {\doibase 10.1016/j.geomphys.2015.05.008} {\bibfield
  {journal} {\bibinfo  {journal} {Journal of Geometry and Physics}\ }\textbf
  {\bibinfo {volume} {96}},\ \bibinfo {pages} {42 } (\bibinfo {year} {2015})},\
  \Eprint {http://arxiv.org/abs/arXiv:1502.04683 [math-ph]} {arXiv:1502.04683
  [math-ph]} \BibitemShut {NoStop}%
\bibitem [{\citenamefont {Connes}(1996)}]{ConnesAction}%
  \BibitemOpen
  \bibfield  {author} {\bibinfo {author} {\bibfnamefont {A.}~\bibnamefont
  {Connes}},\ }\href {\doibase 10.1007/BF02506388} {\bibfield  {journal}
  {\bibinfo  {journal} {Communications in Mathematical Physics}\ }\textbf
  {\bibinfo {volume} {182}},\ \bibinfo {pages} {155} (\bibinfo {year}
  {1996})},\ \Eprint {http://arxiv.org/abs/arXiv:hep-th/9603053}
  {arXiv:hep-th/9603053} \BibitemShut {NoStop}%
\bibitem [{\citenamefont {Connes}\ and\ \citenamefont
  {Lott}(1991)}]{ConnesLott}%
  \BibitemOpen
  \bibfield  {author} {\bibinfo {author} {\bibfnamefont {A.}~\bibnamefont
  {Connes}}\ and\ \bibinfo {author} {\bibfnamefont {J.}~\bibnamefont {Lott}},\
  }\href {\doibase 10.1016/0920-5632(91)90120-4} {\bibfield  {journal}
  {\bibinfo  {journal} {Nuclear Physics B-Proceedings Supplements}\ }\textbf
  {\bibinfo {volume} {18}},\ \bibinfo {pages} {29} (\bibinfo {year}
  {1991})}\BibitemShut {NoStop}%
\bibitem [{\citenamefont {Barrett}(2007)}]{Barrett}%
  \BibitemOpen
  \bibfield  {author} {\bibinfo {author} {\bibfnamefont {J.}~\bibnamefont
  {Barrett}},\ }\href {\doibase 10.1063/1.2408400} {\bibfield  {journal}
  {\bibinfo  {journal} {Journal of Mathematical Physics}\ }\textbf {\bibinfo
  {volume} {48}},\ \bibinfo {pages} {012303} (\bibinfo {year} {2007})},\
  \Eprint {http://arxiv.org/abs/arXiv:hep-th/0608221} {arXiv:hep-th/0608221}
  \BibitemShut {NoStop}%
\bibitem [{\citenamefont {van~den Dungen}(2016)}]{KoenAction}%
  \BibitemOpen
  \bibfield  {author} {\bibinfo {author} {\bibfnamefont {K.}~\bibnamefont
  {van~den Dungen}},\ }\href {\doibase 10.1007/s11040-016-9207-z} {\bibfield
  {journal} {\bibinfo  {journal} {Mathematical Physics, Analysis and Geometry}\
  }\textbf {\bibinfo {volume} {19}},\ \bibinfo {pages} {4} (\bibinfo {year}
  {2016})},\ \Eprint {http://arxiv.org/abs/arXiv:1505.01939 [math-ph]}
  {arXiv:1505.01939 [math-ph]} \BibitemShut {NoStop}%
\bibitem [{\citenamefont {Lizzi}\ \emph {et~al.}(1997)\citenamefont {Lizzi},
  \citenamefont {Mangano}, \citenamefont {Miele},\ and\ \citenamefont
  {Sparano}}]{LizziFermionDoubling}%
  \BibitemOpen
  \bibfield  {author} {\bibinfo {author} {\bibfnamefont {F.}~\bibnamefont
  {Lizzi}}, \bibinfo {author} {\bibfnamefont {G.}~\bibnamefont {Mangano}},
  \bibinfo {author} {\bibfnamefont {G.}~\bibnamefont {Miele}}, \ and\ \bibinfo
  {author} {\bibfnamefont {G.}~\bibnamefont {Sparano}},\ }\href {\doibase
  10.1103/PhysRevD.55.6357} {\bibfield  {journal} {\bibinfo  {journal}
  {Physical Review D}\ }\textbf {\bibinfo {volume} {55}},\ \bibinfo {pages}
  {6357} (\bibinfo {year} {1997})},\ \Eprint
  {http://arxiv.org/abs/arXiv:hep-th/9610035} {arXiv:hep-th/9610035}
  \BibitemShut {NoStop}%
\bibitem [{\citenamefont {Brout}(2001)}]{BroutZitter}%
  \BibitemOpen
  \bibfield  {author} {\bibinfo {author} {\bibfnamefont {R.}~\bibnamefont
  {Brout}},\ }in\ \href {\doibase 10.1142/9789812811585_0016} {\emph {\bibinfo
  {booktitle} {Basics and Highlights in Fundamental Physics: Proceedings of the
  International School of Subnuclear Physics}}},\ \bibinfo {series} {The
  Subnuclear Physics}, Vol.~\bibinfo {volume} {37},\ \bibinfo {editor} {edited
  by\ \bibinfo {editor} {\bibfnamefont {A.}~\bibnamefont {Zichichi}}}\
  (\bibinfo {organization} {World Scientific},\ \bibinfo {year} {2001})\ pp.\
  \bibinfo {pages} {431--468}\BibitemShut {NoStop}%
\bibitem [{\citenamefont {Peres}\ and\ \citenamefont {Terno}(2004)}]{QIandGR}%
  \BibitemOpen
  \bibfield  {author} {\bibinfo {author} {\bibfnamefont {A.}~\bibnamefont
  {Peres}}\ and\ \bibinfo {author} {\bibfnamefont {D.~R.}\ \bibnamefont
  {Terno}},\ }\href {\doibase 10.1103/RevModPhys.76.93} {\bibfield  {journal}
  {\bibinfo  {journal} {Reviews of Modern Physics}\ }\textbf {\bibinfo {volume}
  {76}},\ \bibinfo {pages} {93} (\bibinfo {year} {2004})}\BibitemShut {NoStop}%
\bibitem [{\citenamefont {Somaroo}\ \emph {et~al.}(1999)\citenamefont
  {Somaroo}, \citenamefont {Tseng}, \citenamefont {Havel}, \citenamefont
  {Laflamme},\ and\ \citenamefont {Cory}}]{QuantumSimul1999}%
  \BibitemOpen
  \bibfield  {author} {\bibinfo {author} {\bibfnamefont {S.}~\bibnamefont
  {Somaroo}}, \bibinfo {author} {\bibfnamefont {C.~H.}\ \bibnamefont {Tseng}},
  \bibinfo {author} {\bibfnamefont {T.~F.}\ \bibnamefont {Havel}}, \bibinfo
  {author} {\bibfnamefont {R.}~\bibnamefont {Laflamme}}, \ and\ \bibinfo
  {author} {\bibfnamefont {D.~G.}\ \bibnamefont {Cory}},\ }\href {\doibase
  10.1103/PhysRevLett.82.5381} {\bibfield  {journal} {\bibinfo  {journal}
  {Physical Review Letters}\ }\textbf {\bibinfo {volume} {82}},\ \bibinfo
  {pages} {5381} (\bibinfo {year} {1999})},\ \Eprint
  {http://arxiv.org/abs/arXiv:quant-ph/9905045} {arXiv:quant-ph/9905045}
  \BibitemShut {NoStop}%
\bibitem [{\citenamefont {Eckstein}\ and\ \citenamefont
  {Miller}(2016)}]{2NEW2016}%
  \BibitemOpen
  \bibfield  {author} {\bibinfo {author} {\bibfnamefont {M.}~\bibnamefont
  {Eckstein}}\ and\ \bibinfo {author} {\bibfnamefont {T.}~\bibnamefont
  {Miller}},\ }\href@noop {} {\enquote {\bibinfo {title} {Causal evolution of
  wave packets},}\ } (\bibinfo {year} {2016}),\ \bibinfo {note} {preprint
  \href{http://arxiv.org/abs/1610.00764}{arXiv:1610.00764
  [quant-ph]}}\BibitemShut {NoStop}%
\bibitem [{\citenamefont {Luri\'e}\ and\ \citenamefont
  {Cremer}(1970)}]{DiracSimulSupercond}%
  \BibitemOpen
  \bibfield  {author} {\bibinfo {author} {\bibfnamefont {D.}~\bibnamefont
  {Luri\'e}}\ and\ \bibinfo {author} {\bibfnamefont {S.}~\bibnamefont
  {Cremer}},\ }\href {\doibase 10.1016/0031-8914(70)90004-2} {\bibfield
  {journal} {\bibinfo  {journal} {Physica}\ }\textbf {\bibinfo {volume} {50}},\
  \bibinfo {pages} {224 } (\bibinfo {year} {1970})}\BibitemShut {NoStop}%
\bibitem [{\citenamefont {Schliemann}\ \emph {et~al.}(2005)\citenamefont
  {Schliemann}, \citenamefont {Loss},\ and\ \citenamefont
  {Westervelt}}]{DiracSimulSemicond}%
  \BibitemOpen
  \bibfield  {author} {\bibinfo {author} {\bibfnamefont {J.}~\bibnamefont
  {Schliemann}}, \bibinfo {author} {\bibfnamefont {D.}~\bibnamefont {Loss}}, \
  and\ \bibinfo {author} {\bibfnamefont {R.~M.}\ \bibnamefont {Westervelt}},\
  }\href {\doibase 10.1103/PhysRevLett.94.206801} {\bibfield  {journal}
  {\bibinfo  {journal} {Phys. Rev. Lett.}\ }\textbf {\bibinfo {volume} {94}},\
  \bibinfo {pages} {206801} (\bibinfo {year} {2005})},\ \Eprint
  {http://arxiv.org/abs/arXiv:cond-mat/0410321} {arXiv:cond-mat/0410321}
  \BibitemShut {NoStop}%
\bibitem [{\citenamefont {Zawadzki}\ and\ \citenamefont
  {Rusin}(2011)}]{DiracSimulSemicond2}%
  \BibitemOpen
  \bibfield  {author} {\bibinfo {author} {\bibfnamefont {W.}~\bibnamefont
  {Zawadzki}}\ and\ \bibinfo {author} {\bibfnamefont {T.~M.}\ \bibnamefont
  {Rusin}},\ }\href {\doibase 10.1088/0953-8984/23/14/143201} {\bibfield
  {journal} {\bibinfo  {journal} {Journal of Physics: Condensed Matter}\
  }\textbf {\bibinfo {volume} {23}},\ \bibinfo {pages} {143201} (\bibinfo
  {year} {2011})},\ \Eprint {http://arxiv.org/abs/arXiv:1101.0623
  [cond-mat.mes-hall]} {arXiv:1101.0623 [cond-mat.mes-hall]} \BibitemShut
  {NoStop}%
\bibitem [{\citenamefont {Katsnelson}\ \emph {et~al.}(2006)\citenamefont
  {Katsnelson}, \citenamefont {Novoselov},\ and\ \citenamefont
  {Geim}}]{DiracSimulGraphene}%
  \BibitemOpen
  \bibfield  {author} {\bibinfo {author} {\bibfnamefont {M.~I.}\ \bibnamefont
  {Katsnelson}}, \bibinfo {author} {\bibfnamefont {K.~S.}\ \bibnamefont
  {Novoselov}}, \ and\ \bibinfo {author} {\bibfnamefont {A.~K.}\ \bibnamefont
  {Geim}},\ }\href {\doibase 10.1038/nphys384} {\bibfield  {journal} {\bibinfo
  {journal} {Nature Physics}\ }\textbf {\bibinfo {volume} {2}},\ \bibinfo
  {pages} {620} (\bibinfo {year} {2006})},\ \Eprint
  {http://arxiv.org/abs/arXiv:cond-mat/0604323} {arXiv:cond-mat/0604323}
  \BibitemShut {NoStop}%
\bibitem [{\citenamefont {Lamata}\ \emph {et~al.}(2007)\citenamefont {Lamata},
  \citenamefont {Le\'on}, \citenamefont {Sch\"atz},\ and\ \citenamefont
  {Solano}}]{QuantumSimulDirac}%
  \BibitemOpen
  \bibfield  {author} {\bibinfo {author} {\bibfnamefont {L.}~\bibnamefont
  {Lamata}}, \bibinfo {author} {\bibfnamefont {J.}~\bibnamefont {Le\'on}},
  \bibinfo {author} {\bibfnamefont {T.}~\bibnamefont {Sch\"atz}}, \ and\
  \bibinfo {author} {\bibfnamefont {E.}~\bibnamefont {Solano}},\ }\href
  {\doibase 10.1103/PhysRevLett.98.253005} {\bibfield  {journal} {\bibinfo
  {journal} {Physical Review Letters}\ }\textbf {\bibinfo {volume} {98}},\
  \bibinfo {pages} {253005} (\bibinfo {year} {2007})},\ \Eprint
  {http://arxiv.org/abs/arXiv:quant-ph/0701208} {arXiv:quant-ph/0701208}
  \BibitemShut {NoStop}%
\bibitem [{\citenamefont {Rusin}\ and\ \citenamefont
  {Zawadzki}(2010)}]{ZitterMagnetic}%
  \BibitemOpen
  \bibfield  {author} {\bibinfo {author} {\bibfnamefont {T.~M.}\ \bibnamefont
  {Rusin}}\ and\ \bibinfo {author} {\bibfnamefont {W.}~\bibnamefont
  {Zawadzki}},\ }\href {\doibase 10.1103/PhysRevD.82.125031} {\bibfield
  {journal} {\bibinfo  {journal} {Physical Review D}\ }\textbf {\bibinfo
  {volume} {82}},\ \bibinfo {pages} {125031} (\bibinfo {year} {2010})},\
  \Eprint {http://arxiv.org/abs/arXiv:1008.1428 [quant-ph]} {arXiv:1008.1428
  [quant-ph]} \BibitemShut {NoStop}%
\bibitem [{\citenamefont {Eckstein}\ and\ \citenamefont
  {Miller}(2015)}]{EcksteinMiller2015}%
  \BibitemOpen
  \bibfield  {author} {\bibinfo {author} {\bibfnamefont {M.}~\bibnamefont
  {Eckstein}}\ and\ \bibinfo {author} {\bibfnamefont {T.}~\bibnamefont
  {Miller}},\ }\href@noop {} {\enquote {\bibinfo {title} {Causality for
  nonlocal phenomena},}\ } (\bibinfo {year} {2015}),\ \bibinfo {note} {preprint
  \href{http://arxiv.org/abs/1510.06386}{arXiv:1510.06386
  [math-ph]}}\BibitemShut {NoStop}%
\bibitem [{\citenamefont {Streater}\ and\ \citenamefont
  {Wightman}(2000)}]{Wightman}%
  \BibitemOpen
  \bibfield  {author} {\bibinfo {author} {\bibfnamefont {R.~F.}\ \bibnamefont
  {Streater}}\ and\ \bibinfo {author} {\bibfnamefont {A.~S.}\ \bibnamefont
  {Wightman}},\ }\href@noop {} {\emph {\bibinfo {title} {{PCT}, {S}pin and
  {S}tatistics, and {A}ll {T}hat}}},\ Princeton Landmarks in Mathematics and
  Physics\ (\bibinfo  {publisher} {Princeton University Press},\ \bibinfo
  {year} {2000})\BibitemShut {NoStop}%
\bibitem [{\citenamefont {Franco}\ and\ \citenamefont
  {Wallet}(2016)}]{causMoyal}%
  \BibitemOpen
  \bibfield  {author} {\bibinfo {author} {\bibfnamefont {N.}~\bibnamefont
  {Franco}}\ and\ \bibinfo {author} {\bibfnamefont {J.-C.}\ \bibnamefont
  {Wallet}},\ }in\ \href@noop {} {\emph {\bibinfo {booktitle} {{N}oncommutative
  {G}eometry and {O}ptimal {T}ransport}}},\ \bibinfo {series} {{C}ontemporary
  {M}athematics}, Vol.\ \bibinfo {volume} {676}\ (\bibinfo  {publisher}
  {{A}merican {M}athematical {S}ociety},\ \bibinfo {year} {2016})\ \Eprint
  {http://arxiv.org/abs/arXiv:1507.06559 [math-ph]} {arXiv:1507.06559
  [math-ph]} \BibitemShut {NoStop}%
\bibitem [{\citenamefont {Amelino-Camelia}\ \emph {et~al.}(1998)\citenamefont
  {Amelino-Camelia}, \citenamefont {Ellis}, \citenamefont {Mavromatos},
  \citenamefont {Nanopoulos},\ and\ \citenamefont {Sarkar}}]{DSR}%
  \BibitemOpen
  \bibfield  {author} {\bibinfo {author} {\bibfnamefont {G.}~\bibnamefont
  {Amelino-Camelia}}, \bibinfo {author} {\bibfnamefont {J.}~\bibnamefont
  {Ellis}}, \bibinfo {author} {\bibfnamefont {N.}~\bibnamefont {Mavromatos}},
  \bibinfo {author} {\bibfnamefont {D.}~\bibnamefont {Nanopoulos}}, \ and\
  \bibinfo {author} {\bibfnamefont {S.}~\bibnamefont {Sarkar}},\ }\href
  {\doibase 10.1038/31647} {\bibfield  {journal} {\bibinfo  {journal} {Nature}\
  }\textbf {\bibinfo {volume} {393}},\ \bibinfo {pages} {763} (\bibinfo {year}
  {1998})},\ \Eprint {http://arxiv.org/abs/arXiv:astro-ph/9712103}
  {arXiv:astro-ph/9712103} \BibitemShut {NoStop}%
\bibitem [{\citenamefont {Brukner}(2014)}]{QuantumCausality}%
  \BibitemOpen
  \bibfield  {author} {\bibinfo {author} {\bibfnamefont {{\v{C}}.}~\bibnamefont
  {Brukner}},\ }\href {\doibase 10.1038/nphys2930} {\bibfield  {journal}
  {\bibinfo  {journal} {Nature Physics}\ }\textbf {\bibinfo {volume} {10}},\
  \bibinfo {pages} {259} (\bibinfo {year} {2014})}\BibitemShut {NoStop}%
\end{thebibliography}%
\end{document}